\documentclass[eqsecnum]{emulateapj}
\usepackage[utf8]{inputenc}
\usepackage{natbib,amsmath}
\usepackage{epsfig}
\usepackage{hyperref}
\usepackage[normalem]{ulem}
\DeclareUnicodeCharacter{02BC}{'}

\shortauthors{Gaur et al.}

\begin{document}

\shorttitle{Hard X-ray emission of 3C 273}

\title{Signature of Seyfert-like component in a blazar 3C 273 and its reflection based explanation }

\author{Haritma Gaur\altaffilmark{1}, Main Pal\altaffilmark{2},  Muhammad S. Anjum\altaffilmark{3}, Kiran Wani\altaffilmark{1,4}, Pankaj Kushwaha\altaffilmark{5}, Ashwani Pandey\altaffilmark{6},  Liang Chen\altaffilmark{7}}
\altaffiltext{1}{Aryabhatta Research Institute of Observational Sciences (ARIES), Manora Peak, Nainital -- 263002, India}
\altaffiltext{2}{Department of Physics, Sri Venkateswara College, University of Delhi, Benito Juarez Road, Dhaula Kuan,  New Delhi -- 110021, India}
\altaffiltext{3}{College of Physics and Electronic Engineering, Nanyang Normal University Nanyang, Henan, 473061, Peopleʼs Republic of China}
\altaffiltext{4}{School of Physical Sciences, SRTM University, Nanded, 431606, India}
\altaffiltext{5}{Department of Physical Sciences, Indian Institute of Science Education and Research Mohali, Knowledge City, Sector 81, SAS Nagar,
Punjab 140306, India}
\altaffiltext{6}{Center for Theoretical Physics, Polish Academy of Sciences, Al. Lotników 32/46, 02-668 Warsaw, Poland }
\altaffiltext{7}{Shanghai Astronomical Observatory, Chinese Academy of Sciences, 80 Nandan Road, Shanghai 200030}
\altaffiltext{8}{Corresponding authors Haritma Gaur harry.gaur31@gmail.com, Main Pal rajanmainpal@gmail.com, Liang Chen
chenliang@shao.ac.cn}

\begin{abstract}
 We present the results of blazar 3C 273 obtained from simultaneous observations obtained using XMM--Newton and NuSTAR satellites 
during the period 2015--2019 in five epochs. When the spectra are modeled with a power-law, significant residuals arise below $\sim 2$ keV and in the energy range of 30--78 keV in NuSTAR data. Residuals in the lower energy band represent soft X-ray 
excess while at higher energies it likely represents Compton reflection hump which might be a weak component arising from dense and cold material. The presence of a faint iron line is present in XMM--Newton observations. We interpret such features as attributed to the coronal emission plus the arising from reflection off an accretion disk. We model the SEDs with the single zone 
inverse Compton jet model based on Synchrotron Self Compton and External Compton phenomena. It is found that a one-zone 
synchrotron+IC model explains quite well the SEDs but the jet component alone fails to fit the multiband X-ray emission for the 
low state of this object in 2018 and 2019 which arises due to spectral flattening at low energy X-rays, indicating that an additional Seyfert-like thermal component must be present at X-rays. This is further supported by a big blue bump present in the optical/ultraviolet band in all SEDs. Finally, we analyzed all the epochs using relxill model to incorporate relativistic reflection to model those residuals of soft excess and Compton hump in the X-ray bands. 

\end{abstract}


\section{Introduction}

3C 273 is the nearest high-luminosity quasar with a redshift of z=0.158 and is classified as radio-loud. It is categorized as a Flat Spectrum Radio Quasar (FSRQ) and exhibits many characteristics typical of blazars, including variability in flux across all wavelengths, from radio to gamma-ray energies, as well as variable polarization from the radio to the X-ray spectrum (i.e. \cite{1998A&ARv...9....1C,2008A&A...486..411S}). The quasar features a relativistic jet that demonstrates superluminal motion \citep{2001ApJ...556..738J, 2005AJ....130.1418J}. The emissions detected from radio to optical wavelengths are attributed to synchrotron radiation produced by the jet \citep{2010ApJ...714L..73A}. Additionally, the presence of a prominent feature known as the Big Blue Bump (BBB) in the optical/UV spectrum can be accounted for by two independently varying components; the more rapidly varying component is believed to arise from the reprocessing of photons in an accretion disk, while the slowly varying component may be associated with jet emissions \citep{1998A&A...340...47P}.\\
\\
In the low-energy X-ray band ($<$2 KeV),this source exhibits a soft X-ray excess (SE) \citep{2008A&A...479..365P}. This phenomenon may be attributed to the blue tail of the multi-colour black body components or due to the Comptonization of UV photons \citep{2004MNRAS.349...57P}. 
Alternatively, it could arise from significant smearing of reflected emissions arising from the accretion disk very close to the 
supermassive black hole (SMBH) \citep{2006MNRAS.365.1067C}. The comptonization process is favoured due to the observed correlated variations between SE and optical/UV emissions \citep{1992A&A...258..255W, 2004MNRAS.349...57P}; however, 
later \cite{2007A&A...465..147C} could not find any substantial correlation between these bands. Using simultaneous observations conducted with XMM--Newton, \cite{2017MNRAS.469.3824K} discovered a strong correlation between UV and X-ray emissions during the low state of 3C 273. In contrast, during the flaring state, the relationship between optical/UV and X-ray emissions exhibits different behavior. \\
\\
In previous studies, a faint Fe-K$\alpha$ line has been detected around 6.4 keV in the X-ray spectrum of this source 
\citep{1990MNRAS.244..310T, 2004Sci...306..998G, 2002MNRAS.336..932K,2015ApJ...812...14M, 2017MNRAS.469.3824K}. Based on these 
characteristics, it has been inferred that the emission of this blazar in X-ray bands results from a mixture of thermal 
processes specifically, the component originating from the inner accretion disk possibly from the corona 
that might be accompanied by disk reflection and non-thermal processes are associated with the jet emission.\\
\\ 
Previous X-ray observations of this source between 2 to 10 keV were well-fitted with a powerlaw having spectral 
index of $\sim$1.5 \citep{1985SSRv...40..623T, 2007A&A...465..147C}. NuSTAR observations of \cite{2015ApJ...812...14M} found that the spectrum of this object is well described by a photon index of $\sim$1.5 upto 3-20 keV energy band. 
% whereas the 2---25 keV band can be well fitted with a powerlaw of 
%$\Gamma$=0.91 \cite{1985SSRv...40..623T} which is common for jet dominated blazars. 
%The powerlaw spectrum can be extended up to 200 keV by BeppoSAX data with $\Gamma$ $\sim$1.6 \citep{1998A&A...340...35H}.
In many previous studies, deviation from powerlaw model is found speciafically below 2 keV and above 10 keV  
 (i.e. \citep{2015ApJ...812...14M,2004Sci...306..998G} which required a reflection model including a cold reflector in addition
 to powerlaw such as PEXRAV. \cite{2007A&A...465..147C} used the broadband spectrum from 0.2--100 keV made with quasi-simultaneous XMM--Newton and INTEGRAL data to find an explanation for the disk reflection hump at around 30 keV. \cite{2015ApJ...812...14M} found that the X-ray spectrum obtained from NuSTAR satellite showed significant deviation above 20 keV which can be well analyzed by 
an exponentially cutoff powerlaw which indicates a weak reflection possibly from a dense and cold material with $\Gamma$ $\sim$1.5. 
However, adding INTEGRAL data, they found that an additional powerlaw is required to fit the jet component with $\Gamma$ $\sim$1.05. 
\cite{2015A&A...576A.122E} also studied the broadband spectra of this source which indicated that the X-ray radiation is generated 
likely from the inner disk/reprocessed emission from the corona whereas the $\gamma$-ray emission is primarily dominated by 
emissions from the relativistic jet.\\
\\
The spectral features observed in the spectra of 3C 273, including Fe K$\alpha$, a soft X-ray excess, and an X-ray reflection bump near 30 keV, exhibit variability across different observations, probably dependent on the state of the source \citep{1998A&A...340...35H}. These spectral features are similar to those identified in Seyfert galaxies, which are supposed to have low-power active galactic nuclei (AGNs) in contrast to blazars \citep{1995PASP..107..803U}. Consequently, 3C 273 is a special source as it possesses Seyfert-like components that are probably depends on the flux level of the source. During the high flux state, the jet emission predominates over thermal/reprocessed emissions associated with the accretion disk, resulting in the absence of such features in the spectra. While in a low state, the contribution from thermal is comparable to the non-thermal component and hence the X-ray features become visible in the spectrum. Hence, it is essential to observe the spectra of such sources in their different flux states.\\
\\
In this research work, we studied the contemporaneous observations of 3C 273 using the European Photon Imaging Camera (EPIC)-pn, Optical Monitor telescope aboard XMM--Newton satellite, and NuSTAR observations during the period 2015--2019 in five epochs. The advantage of using simultaneous observations of XMM--Newton and NuSTAR provide an opportunity to probe the entire energy range from 
0.4-78 keV to search for any signature of thermal/non-jetted components in the X-ray emission. The presence of soft X-ray excess, 
the presence of weak iron line emission, and the Compton reflection hump peaking near 20 keV are generally considered to be prominent feature of an AGN \citep{2002MNRAS.336..932K, 2004Sci...306..998G}. \cite{2015ApJ...812...14M} studied this blazar during 2012 and found a weak reflection component in the X-ray emission using simultaneous observations from  XMM--Newton and NuSTAR satellite.\\
\\
Motivated by such studies, we aim to search for spectral features that could indicate the presence of thermal/non-jetted components in the X-ray emission of this source using five-year span of observations. We model such features using coronal models which are used to describe the X-ray emission of Seyfert galaxies. The broadband spectral energy distribution (SED) of 3C 273 represents two distinct broad bumps similar to a blazar. We used the quasi-simultaneous multi-wavelength radio to $\gamma$-ray observations to construct its SEDs to discern the big blue bump in optical/UV bands which further enhances the possibility of disk component in its emission. Also, we model the high energy component of SED i.e. X-ray and $\gamma$-rays using the two-component inverse Compton model to check whether the above-described features can be described with the Synchrotron Self Compton (SSC) and the External Compton (EC) models as claimed by \citet{2015ApJ...812...14M}. The data sets and data analysis are presented in the next section. 
The results can be found in section 3, while section 4 contains the discussion. Lastly, the conclusions are outlined in section 5.\\
\\
\section{Observations and Data Reduction}
\subsection{XMM--Newton (EPIC/pn)}

In this study, we analyzed EPIC pn data from XMM-Newton satellite as it is least affected by the effects of photon pile-up. 
The processing and analysis of the X-ray data are performed using the analysis procedures described in \cite{2013Icar..226..186S}.
 We extracted only the single-pixel events for our analysis. Particle/solar flares from the light curve are excluded by removing those periods having count rates higher than 0.4 s$^{-1}$. %The light curve for $E >$ 10 keV 
A circular source region with a radius ranging between 35--40 arsec centered on the source co-ordinates is extracted to prepare the X-ray spectrum of each data set. A circular region having the same radius which is offset by about 180 arcsec from the source is also extracted to prepare the background spectrum. Pile-up effects in each observation is determined by the Science Analysis System (SAS) task EPATPLOT. We did not find any pile-up effects in our observations. We grouped the spectra to have atleast 30 counts in each bin. %We consider only the energy range of 0.4 to 10 keV to eliminate uncertainties associated with calibration below 0.4 keV.
The tasks {\it rmfgen} and {\it arfgen} from SAS were employed to generate the photon redistribution matrix and the ancillary files,
 respectively.\\

\subsection{Optical/UV data}

A 30 cm  Ritchey-Chretien Optical/UV Monitor Telescope (hereafter referred as OM) on board XMM-Newton enables simulateneous observations from optical/ultraviolet to X-ray bands. It provides high imaging sensitivity featuring a focal length of 3.8 m and time resolution of 0.5 s. It operates within a wavelength range of 170 to 650 nm, utilizing six broad-band filters, three for optical observations and three for UV band. The optical filters correspond to U, B, and V bands, operating in the wavelength window of 300 to 390, 390 to 490 and 510 to 580 nm, respectively. The ultraviolet filters correspond to UVW2, UVM2, and UVW1 bands, collecting data in the wavelength ranges of 180 to 225, 205 to 245, and 245 to 320 nm, respectively. During five observational sessions, 3C 273 was monitored by the OM in imaging mode across all filters. The imaging data were reprocessed using the standarad omichain pipeline of the Science Analyisis System (SAS), from which we get a combolistfile that contains calibrated data for all sources within the field of view. The final source list file provides the count rates, instrumental magnitude, and fluxes for 3C 273 which are provided in Table 2.

\subsection{NuSTAR}

We have analyzed {\it NuSTAR} data utilizing HEAsoft version 6.26 along with the latest CALDB version 20191219. {\it NuSTAR} light 
curves and spectra for the blazar 3C 273 were obtained from a circular region of radius 40 arcsecs, using the standard  {\it nupipeline} and {\it nuproducts} scripts. Background data were also obtained from a circular region of the same size. {\it NuSTAR} spectra for 3C 273 were binned to ensure atleast 25 counts per bin utilizing the {\it grppha} tool.

\subsection{Fermi data}

We used the latest Fermi-LAT data processed with the latest PASS 8 (P8R3$_{V2}$) instrument response function (IRF) available at the time of analysis around the XMM-observation duration and followed the standard analysis procedure as prescribed by the LAT team to extract relevant physical quantities. The SED extraction (described below in detail) was performed over a data binning timescale of 1-day, a week, 2-week, a month, and six-month and found that the source is very weak above a few GeVs on all timescales and also the SED points in different bins were within the 1-sigma uncertainty and thus, we used six-month data around each XMM ID.

We used the FERMITool (v 1.0.1) analysis package for the analysis. To prepare the data files for spectral analysis, we first selected events tagged as "SOURCE" class (evclass=128) with both front + back events (evtype=3) and energies between 0.1 to 300 GeV from a circular region of 15 degrees centered on the source. During this selection, a zenith angle constraint of 90 degrees was also applied to avoid gamma-ray photons from the Earth's limb. Following this, we generated good time intervals using the standard flag ``(DATA\_QUAL$>$0)\&\&(LAT\_CONFIG==1)''. Then, we generated the exposure map on the source region (15 degrees) plus an additional annular region of 10 degrees around it using the task "gtexpomap" required for the spectral analysis. 

For the spectral fitting, we used the "unbinned maximum likelihood" analysis method "pyLikelihood" provided with the analysis package. In addition to the event and exposure file, it requires an XML spectral model file which we generated from the 3rd LAT point source catalog (3FGL; \cite{2015ApJS..218...23A} ) with the source spectrum model as power-law. The spectral fitting was performed iteratively, by removing sources with test statistics (TS) $<$ 1 and fixing the spectral indices of sources with TS $\le$ 9 as well as the successive higher TS source until the fit converged. 

The SED extraction was done in two steps. First, we fitted the whole 0.1 -- 300 GeV band and used the XML model file generated from
this for the SED extraction. We used six various energy bins: 0.1 -- 0.3, 0.3 -- 1, 1 -- 3, 3 -- 10, 10 -- 100, and 100 -- 300 GeV for the SED construction using a power-law model for each bin. The above-mentioned procedure selection and spectral fitting
were repeated until the fit converged for each energy bin. For scientific analysis, only bins with TS $\ge$ 9 (~ 3-sigma detection criteria) were considered.

\subsection{Radio data}

Crimean Astronomical Observatory (CrAO) conducted simultaneous radio observations at a frequency of  22.2 GHz 
with the 22 m radio telescope (RT-22). For the measurements at 22.2 and 36.8 GHz, two identical Dicke-switched radiometers were used. The position of the source was determined with scanning. The radio telescope was pointed towards the source using the principal and reference beam lobes, which were created during beam modulation and having mutually orthogonal polarizations. A series of 6 to 20 measurements were taken depending upon the source;s intensity, after which the mean signal intensity was claculated. The 14 m radio telescope at Aalto University’s Metsahovi Radio Observatory in Finland was utilized for observations at 37.0 GHz. The data colltected from RT22 and Metsahovi were combined to supplement each other. Detailed description of the data reduction process can be seen in \cite{1998A&AS..132..305T} and \cite{2015A&A...582A.103G}.

\begin{figure}
\centering
\includegraphics[width=9cm , angle=0]{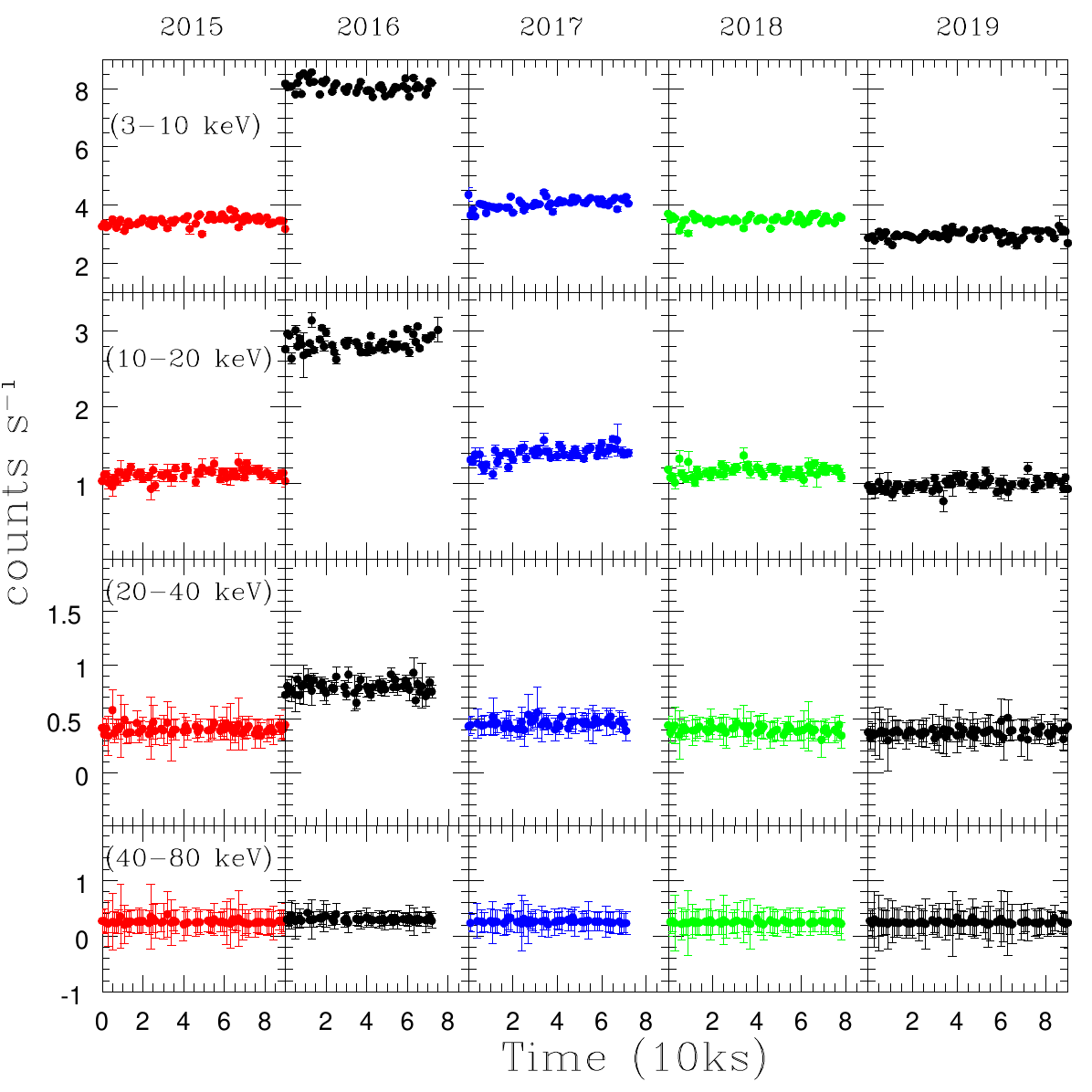}
\caption{Light curve of NuSTAR in different energy bands for 3C 273. }
\end{figure}

\begin{figure}
\centering
\includegraphics[width=7cm , angle=0]{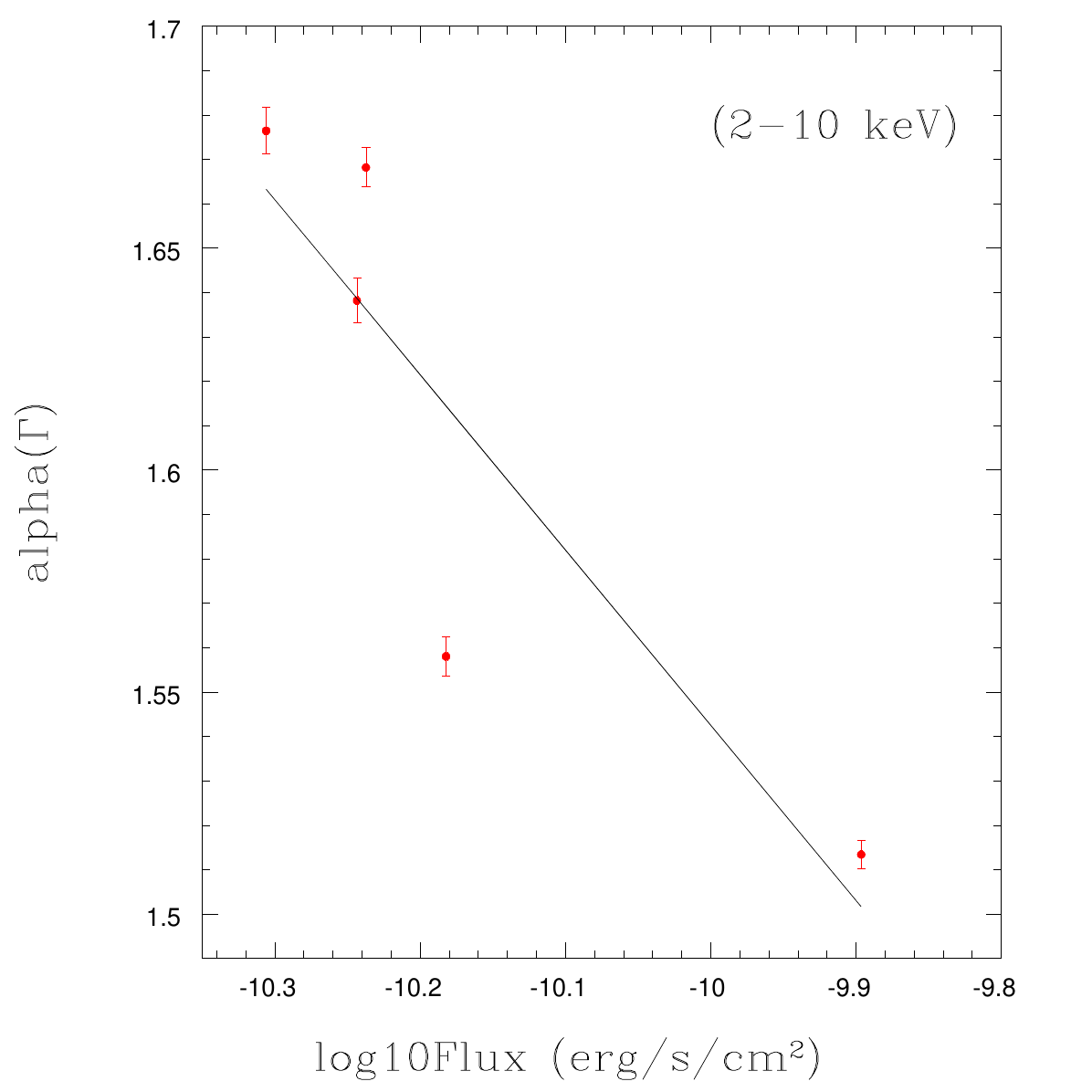}
\caption{the 2-10 keV Flux and powerlaw spectral index $\Gamma$ in the energy range 2--10 keV. }
 \end{figure}

\section{Results}
 \subsection{Variability and spectral evolution}

To find out the flux changes during these five years from 2015--2019, we estimated the noise subtracted variance fraction of the 
mean i.e. $F_{var}$  \citep{2002ApJ...568..610E}. More details are provided by \cite{2017ApJ...850..209G}. We found the variability
 amplitude of 0.75--3 \% in the XMM--Newton and NuSTAR light curves for this source. \cite{2017MNRAS.469.3824K} also found moderate flux variability in the X-ray light curves of this source with a maximum fractional variability of 2.6\%. To investigate $F_{var}$ as a function of energy during these observations, we decomposed the NuSTAR light curves into four energy bands i.e. 3 to 10, 10 to 20, 20 to 40, and 40 to 78 keV (shown in Fig 1.) and calculated variability amplitude in all these energy bands. We did not find significant variability amplitude at higher energies.\\
\\
Most of the previous observations pointed to the existence of the "coronal/Seyfert-like component" (presence of soft excess, weak iron line) in the broadband X-ray spectra, along with the non-thermal jet component \citep{1998PASJ...50..213C,2002MNRAS.336..932K,2004MNRAS.349...57P,2007A&A...465..147C}. This Seyfert-like component is generally thought to be the underlying unbeamed non-thermal component and should be prominent during the low flux states of the source. An estimate of low flux state as $F_{2-10 keV}$ $\leq$ 10$\times 10^{-11}$ erg $cm^{-2} s^{-1}$ was provided by \cite{1998PASJ...50..213C}. A model was proposed by \cite{1998A&A...340...35H} that incorporates a Seyfert-like feature superimposed over a time-variable beamed jet component, which becomes deteactable when the jet component exhibits a low flux level. Nevertheless, \cite{2002MNRAS.336..932K} employed RXTE data to show that the superposition of a Seyfert-like component on the beamed jet component does not correlate with the flux state. Similar conclusions were later presented by \cite{2004Sci...306..998G}.\\
\\
Here, we investigate the flux state of the source during our observations. \cite{2017MNRAS.469.3824K} studied X-ray spectra of this
 source during 2000-2015 using XMM--Newton observations and compared the flux states of their observations with the previous literature. They reported that the minimum flux state of this source is found in 2015 which is 4.98 $\times10^{-11}$ erg $cm^{-2} s^{-1}$ (flux is calculated in 2.5--10 keV energy band). We also analyzed the XMM--Newton observation of 2015 and found its flux to be 5.79 $\times10^{-11}$ erg $cm^{-2} s^{-1}$ in 2--10 keV energy band. However, we noticed that the source is in the even fainter state in 2019 which is 4.94 $\times10^{-11}$ erg $cm^{-2} s^{-1}$. Hence, we can say that 3C 273 is in its lowest flux state in 2019. 
Comparing all these five observations, 3C 273 is in its brightest state in 2016 having a flux of 12.70 $\times10^{-11}$ erg $cm^{-2} s^{-1}$ which is lower as compared to flux in its flaring period during 2007 \citep{2007A&A...465..147C, 2015A&A...576A.122E}. Hence, we can say that the XMM--Newton observations analyzed in our work are performed mostly during the low flux state of 3C 273 except for observation performed in 2016.\\
 \\
The spectral variation of 3C 273 is shown in Figure 2  and it can be seen that $\Gamma$ varies between 1.51--1.68 during 2015--2019. We fitted a linear function and found that a "harder-when-brighter" trend is found in this source which is consistent with previous studies \citep{2015ApJ...812...14M}. It is claimed that such variation is attributed to contribution of jet emission which increases with flux. The reflection-dominated AGNs such as Seyferts may also exhibit a similar nature when high energy radiation is reflected from the surroundings such as accretion disk and nearby thick cloud gases.\\
\\

\subsection{Spectral Fitting using XMM--Newton and NuSTAR observations}

In earlier studies of 3C 273, it has been determined that the X-ray emission exhibits greater complexity than the simple
interpretation of nonthermal jet emission. A soft excess component is detected in this source \citep{2004MNRAS.349...57P, 2008A&A...479..365P}. A weak iron line emission and X-ray reflection bump at around 30 keV is also detected in this source (\cite{2015ApJ...812...14M} and references therein). Hence, we analyzed simultaneous XMM--Newton and NuSTAR data to achieve extensive and 
high quality spectral coverage ranging from 0.4--78 keV which is useful in probing the presence of soft excess/iron line and 
constraining the X-ray reflection component of this source.\\
\\ 
XSPEC version 12.10.1 software is used for spectral fitting, with all quoted errors representing 2 $\sigma$ (90\%). 
The spectral fitting was conducted over the energy range of 0.4 to 78 keV for all spectra analyzed. The fitting process employed Wilms abundances \citep{2000ApJ...542..914W} and Verner cross-sections \citep{1996ApJ...465..487V}, while the hydrogen column density is fixed at the Galactic value of 1.79 $\times$ $10^{20} cm^{-2}$ \citep{1990ARAA..28..215D}. \\
\\
We fitted all these five observations using the powerlaw model and found that there are deviations from the powerlaw model which 
are generally expected from this source as it is reported to have soft excess, a reflection component with a weak iron line, and a Compton hump \citep{2015ApJ...812...14M, 2008A&A...479..365P}. \\
\\

\subsubsection{Possible Reflection signatures}

{\bf (i) Soft X-ray excess}

It is evident from Figure 3 that the X-ray spectra for all observations of 3C 273 exhibit a soft excess below 2 keV. 
In literature, the existence of a soft excess has been established \citep{1990MNRAS.244..310T, 1998PASJ...50..213C, 
2004Sci...306..998G, 2004MNRAS.349...57P}; however, its origin is not clear. The soft X-ray excess is modeled using black body components as indicated by \cite{2004Sci...306..998G}. In contrast, \cite{2004MNRAS.349...57P} preferred Comptonization models, interpreted the soft X-ray excess arises from the Comptonization of accretion disk photons by secondary thermal electrons of different population having a temperature of $\sim$350 eV. Furthermore, \cite{2008A&A...479..365P} explained the Seyfert-like component through a non-thermal corona model, where soft seed photons from the disk illuminate a corona composed of blobs or loops. In this scenario, non-thermally distributed electrons inverse Comptonize the soft seed photons, resulting in the coronal component of the X-ray spectrum. Our analysis showed a significant soft X-ray excess across all five observations presented, as illustrated in Figure 3.\\
\\ 
\noindent
{\bf (ii) Fe K$\alpha$ emission}

The iron line detection at 6.4 keV in 3C 273 has been reported several times in the past but most of the time, it remains
faint or even absent. The detection of the Fe K$\alpha$ line serves as an indicator of X-ray reprocessing by cold material, suggesting that both thermal and non-thermal emission processes significantly contribute to the emission characteristics of this specific blazar. Fe--K$\alpha$ emission line results from the fluorescence phenomenon after photo-absorption of high energy photons in the accretion disc material or distant material such as torus. Sometimes this Fe--K$\alpha$ emission line is too broad to be detected due to relativistic effects. The mechanism that govern radiation and its frequency of visibility within the X-ray spectrum are not known completely.\\
\\
Iron line observed in the X-ray spectra is caharecterized by a broad profile ($\sigma$ $\sim$ 0.6 keV, EW $\sim$ 20--60 eV).
This line is sometimes neutral, as noted in previous studies \citep{1990MNRAS.244..310T, 2004Sci...306..998G}, while sometimes it is found to be ionized \citep{2000ApJ...544L..95Y, 2002MNRAS.336..932K}. We checked for the presence of this Fe K$\alpha$ line in our observations and the spectral fitting results to a powerlaw plus Gaussian profile model are provided in Table 4. We add a Gaussian profile to a powerlaw model;
 fixing the peak energy and allowing line width to vary (following \cite{2000ApJ...544L..95Y}), we calculated the best fitting parameters which are presented in Table 3. $\sigma$ varies from 0.17--0.41 keV and equivalent width varies from 7.25--20.37 eV. \\
\\

\noindent
{\bf (iii) Compton reflection component}

In order to measure the spectral properties of hard X-ray band, pexrav model (i.e. \cite{1995MNRAS.273..837M}) is generally used, 
which is the addition of a cutoff powerlaw and reflection from a plane parallel slab consists of neutral material. In previous studies, pexrav model has been extensively used, and consequently, we have employed it in our study (within the energy range of 2-78 keV to avoid complications due to soft excess) to facilitate a direct comparison of our findings with those from earlier studies. \\
\\
This pexrav model incorporates the photon index $\Gamma$, the high energy cutoff $E_{c}$, and the reflection component strength R. 
The parameter R quantifies the strength of the reflection with respect to the primary emission. We have adopted solar abundances for the reflector and set the inclination angle to cos i = 0.45, which is the model's default value. During the fitting process, the free parameters are the cross normalization constant between FPMA and FPMB, the reflection scaling factor R, photon index $\Gamma$, and the normalization of the cutoff powerlaw. The parameters of this model enhance the fitting statistics when compared to a powerlaw. We found similar results as quoted by \cite{2015ApJ...812...14M}.\\
\\
\subsection{Broadband SED Modeling}

3C 273 is classified as a blazar, characterized by its relativistic jet that is aligned at a low angle relative to the 
observer's line of sight. This alignment is further indicated by a prominent Big Blue Bump observed in its optical 
spectrum \citep{2006ApJ...653L...5G}. The existence of 
this Big Blue Bump in the optical-UV spectrum may suggest that thermal emissions from the accretion disk are contributing to the 
observed radiation \citep{1999ApJ...527..683K}. The spectral characteristics derived from the X-ray data can 
be explained by a two component inverse Compton emission originating from the jet, as proposed by \cite{2015ApJ...812...14M}. 
In this context, the slight bump detected in the NuSTAR data, which is represented by Compton reflection from the accretion disk, 
could potentially be ascribed to Synchrotron Self-Compton (SSC) emission, as illustrated in references such as 
\cite{1992ApJ...397L...5M}.
The high energy emission (MeV to GeV range) could be attributed to the scattering of the radiation external to the jet 
(EC scenario, see, e.g., \cite{1994ApJ...421..153S}). In this viewpoint, we model the broadband SEDs of 3C 273 corresponding to different epochs in the context of a stationary one-zone model assuming synchrotron and inverse Compton (IC) emission mechanisms \citep{2009MNRAS.397..985G, 2018ApJS..235...39C, 2020ApJ...898...48A}. We compiled the broadband SEDs of 3C 273 for 2015, 2016, 2017, 2018, and 2019 epochs by incorporating simultaneous radio data (at 22.2 and 36.8 GHz), XMM-Newton X-ray and Optical-UV data, Fermi-LAT gamma-ray data to determine the physical parameters of a one-zone blazar model. The archival radio and IR data from the Space Science Data Center (SSDC)\footnote{https://www.ssdc.asi.it} is used to tightly constrain the synchrotron peak of the SED.\\
\\
We assume an accelerated broken power law electron energy distribution (EED) injected in a homogeneous spherical emitting region of size $R$, filled with a random magnetic field $B$. The spherical blob moves with relativistic speed $\beta$ and Lorentz factor $\Gamma=(1-\beta)^{-0.5}$, making a very small angle $\theta$ with the line of sight. The emitting broken power law EED naturally arises due to a competition between acceleration and cooling in the jet and is given between a low energy $\gamma_{min}$ and high energy $\gamma_{max}$ as

\begin{equation}
\centering
N(\gamma)=\left\{
    \begin{array}{ll}
    k\gamma^{-p_{1}}    &\mbox{$\gamma\leq\gamma_{b}$}\\
    k{\gamma_{b}}^{p_{2}-p_{1}}\gamma^{-p_{2}}   &\mbox{$\gamma>\gamma_{b}$},\\
    \end{array}
\right.
\end{equation}

where $ \gamma_b $ is the broken energy, $k$ is the normalization constant, and $p_{1}$ and $p_{2}$ are spectral indices of the EED below and above the break energy. The emission is enhanced by the Doppler factor $\delta$ as given by

\begin{equation}
\centering
\delta = \frac{1}{\Gamma \left( 1 - \beta \cos \theta \right) }.
\end{equation}

We assume $\theta \sim 1/\Gamma$ in our model that leads to $\Gamma=\delta$. The size of the emitting source $R$ is associated with 
the observed variability timescale $\Delta t$ as follows:

\begin{equation}
\centering
R= c \Delta t \delta.
\end{equation}

Blazars usually show a day-scale variability, with the timescale getting shorter at higher frequencies. The blazar emission is highly boosted with Doppler factor $\delta= 5-50$, making the observed variability timescales appear shorter.\\

We compute the synchrotron and IC emission from the stationery broken power law EED injected in the spherical blob. For IC emission, we include the synchrotron self-Compton (SSC) process and the external Compton (EC) scattering on ambient photons using a full Klein-Nishina cross-section. The ratio of observed Compton bump frequency $\nu_c$ and synchrotron peak frequency $\nu_s$ in an IC model is given by

\begin{equation}
\centering
\frac{\nu_c}{\nu_s}= \frac{4}{3} \gamma_b^2
\end{equation}

where $\gamma_b$ is the break energy of the EED. The ratio of Compton to synchrotron power, called Compton dominance, is given as

\begin{equation}
\centering
\frac{\nu_c Fc}{\nu_s F_s}= \frac{U'_{rad}}{U'_B},
\end{equation}

with $U'_B= B^2 /8 \pi$ and $U'_{rad}$ being the magnetic and ambient seed photon energy densities in the jet frame. The seed 
photon contributions for IC emission come from the synchrotron itself in the case of SSC and, in the case of EC, from BLR (taken 
as a blackbody centered on Ly$\alpha$) and a relatively colder dusty torus (taken as a blackbody at a temperature 
$T_{DT}= 10^2$ K). 

For SSC, the synchrotron seed photon energy density in the jet is given by
\begin{equation}
\centering
U_s' = \frac{Ls}{4\pi R^2 c \delta^4}.
\end{equation}

For EC, the energy density $U_{EC}'$ of ambient BLR or dusty torus photons is related to source accretion disk luminosity $L_D$  as given by
\begin{equation}
\centering
U_{EC}' = \frac{17}{12} \Gamma^2 \zeta  \frac{L_D}{4\pi R^2_{EC} c},
\end{equation}
where $R_{EC}$ is the radius of BLR or dusty torus. We assume that BLR and dusty torus reprocess a 10\% of accretion disk emission (i.e., $\zeta= 0.1$).\\
\\
We fit the observed SEDs of 3C 273 for different epochs using a $\chi^2$ minimization procedure. The best model fits to the SED 
data are shown in Figure 4 and the physical parameters are provided in Table 5. It can be seen that the entire broadband SEDs of 
3C 273 are dominated by non-thermal emissions arising from SSC and EC emissions similar to typical blazars. The X-ray emission is 
brighter than $\gamma$-rays, contrary to typical blazars. The GeV emission is dominated by EC emission whereas the X-ray is 
dominated by SSC emission. Both EC components based on BLR and dusty torus (EC-BLR and EC-DT, respectively) are needed to 
adequately fit the Fermi-LAT spectra, suggesting that the location of the emitting blob might be at the edge of BLR, where both 
BLR and dusty torus contribute significantly to observed $\gamma$-ray emission. The average physical jet parameters are $R\approx 10^{17}$ cm, $B\approx 0.4$ G, $\gamma_b \sim 10^3$, and $\delta \approx 6$, that are usually found for typical blazars. A relatively low value of the Doppler factor indicates that emission 3C 273 should not be significantly Doppler boosted and, therefore, any thermal components may not be swamped by the jet. This is in agreement with the fact that all SEDs show a thermal big blue bump \citep[also see, e.g.,][]{2020MNRAS.497.2066F}, fitted with a black body at a temperature $T_D \approx 2\times 10^4$ K with peak luminosity $L_D \approx 2\times 10^{46}$ erg s$^{-1}$.\\
\\
Figure 4 shows that a one-zone model can fit the overall SED of the source self-consistently but seems problematic in explaining the 
multi-band X-ray emission, during the low states of 3C 273 in 2018 and 2019. The multi-band X-ray data for these states 
deviate significantly 
from the expectation of the model due to spectral flattening at low-energy X-rays. The model underestimates the high-energy X-ray 
emission. Any attempt to fit the high-energy X-rays would underestimate the low-energy X-ray emission. This implies that the non-thermal 
jet component is not sufficient to explain the multi-band X-ray emission and an additional thermal component might be necessary. Our X-ray 
spectral analysis of 3C 273 shows a soft excess and weak Iron line, indicating the contribution of a Seyfert component along with the 
non-thermal jet component. Hence, based on SED modeling and X-ray spectral analysis, we suggest that a thermal Seyfert-like component 
might significantly contribute to the observed X-ray emission. The presence of a consistent thermal accretion disk for all epochs supports 
the presence of a thermal component at X-rays.

\subsection{Fitting using relxill model}

It has been established in \cite{2015ApJ...812...14M} that the X-ray emission of 3C 273 is a combination of a Seyfert component, which 
includes both reflection from disk and coronal emission, along with a jet powerlaw component. The study indicated that the jet emission starts to 
dominate at energies exceeding 30–40 keV. Consequently, we employed the relxill plus power law model to fit all observations. 
Relxill is a reflection model, integrating xillver \citep{2011ApJ...731..131G, 2013ApJ...768..146G} and relline 
\citep{2010MNRAS.409.1534D, 2013MNRAS.430.1694D}. This model effectively describes reflection features, including the Fe–K$\alpha$ 
emission line around 6 keV and the Compton reflection hump, which peaks near approximately 30 keV. The Compton reflection hump arises 
due to the backscattering of higher energy photons by the disk. This model computes the reflected emission at each angles and 
radii within the accretion disk \citep{2014ApJ...782...76G}. Further details regarding this model and its various applications can be 
found on the associated web page document. \footnote{\url{http://www.sternwarte.uni-erlangen.de/~dauser/research/relxill/index.html}}.\\
\\
The relxillCp model, developed by Dauser et al. and Garcia et al., assumes that the X-ray source illuminates the accretion disk
 through a nthcomp Comptonization process \citep{2004MNRAS.349.1435M}. This illumination is characterized by a broken emissivity law, expressed as $\epsilon (r) \propto r^{-Index1}$ between $R_{in}$ and $R_{br}$, and $\epsilon(r) \propto r^{-Index2}$ for the 
area between $R_{br}$ and $R_{out}$. Here, $r$ represents the radius of the accretion disk, while Index1 and Index2 denote the
inner and outer emissivity profile indices, respectively. The parameters $R_{in}$, $R_{out}$, and $R_{br}$ correspond to the inner, outer, and break radii of the disk. Additional parameters include the inclination angle $(i)$, iron abundance ($A_{Fe}$) relative to solar levels, spin $(a)$, the illuminating power law index $\Gamma$, the ionization parameter $(\xi)$ ($=L_{X}/nr^{2}$), where $L_X$ is the X-ray luminosity of the source and $n$ is the hydrogen number density of the disk material), as well as the reflected fraction ($R$). For the fitting process, the iron abundance is fixed to 1, the inclination angle is set to 35 degrees, the outer radius is fixed at 400 $r_{g}$, the spin parameter is specified as $a=0.998$, and the relative refraction is fixed to unity. All other parameters are allowed to vary during the fitting. %While calculating the error on the parameters, a few parameters were required to be fixed at their best-fit value as shown in Table 4. 
The fit resulted in best  $\chi^{2}$ values as compared to previous models listed in Table 4 and their respective plots are shown in the bottom panel of Fig 3. The accretion disk/corona photon index is steeper (varying from $\Gamma$=1.8--2.0) as compared to the jet photon index (1.1--1.4).  \\

\section{Discussion }

In our studies, we analyzed simultaneous observations from XMM-Newton and NuSTAR of this blazar over the period from 2015 to 2019 in five distinct epochs of observation. During our observations, 3C 273 was mostly in the low state except in 2016, where it is found in a comparatively higher state. We model the X-ray spectra using absorbed power law but the spectra show the soft excess components below 2 keV energy. This indicates the prominence of the Seyfert-like components in the spectra. Therefore, to search for other Seyfert-like components, we model the spectra of XMM--Newton observations using power law plus Gaussian component to infer the presence of faint Fe K$\alpha$ line at 6.4 keV. We found that the quality of fit improves with its inclusion. We found a slight hump at around 30 keV which suggests the presence of a weak reflection continuum which is attributed to the reflection from an accretion disc \citep{2015ApJ...812...14M}. All these evidences infer the presence of a Seyfert-like component in this source. The overall spectral energy distribution of 3C 273 clearly show the presence of the big blue bump which further indicates the dominance of the accretion disk contribution in the overall emission of the source.\\
\\
The multi-band X-ray SED of 3C 273 seems complex and the composite emission appears to arise as a combination of a jet component and a coronal/Seyfert-like component that is the emission from a dynamic corona. Both these components are variable in time but are not correlated as they have different origin. Most of the time jet component dominates or outshines the Seyfert-like component. However, depending on the flux level of the source, a Seyfert-like component can have a significant contribution to the overall X-ray emission. Therefore, it is necessary to model the X-ray spectra of such source during different flux states to discern the contribution of this component. \\
\\
The entire SED extending from radio to gamma-ray wavelengths is dominated by the non-thermal radiation originating from the jet, which is well characterized by SSC and EC processes. However, the existence of thermal emission from the accretion disk has been supported by observations of a soft excess in the X-ray spectrum, presence of a weak iron line, a prominent blue bump in the optical/UV range, and a Compton reflection hump. These features are typically associated with an AGN as noted in previous studies (e.g., \cite{2002MNRAS.336..932K, 2008A&A...479..365P}). All of the above features or a few of them might be present in the spectra and they may vary depending on the source's flux state, which could obscure Seyfert-like signatures due to the significant contribution of high jet flux to the overall emission of the blazar. \\
\\
In previous research, the authors found evidence of AGN components within their observations and tried to elucidate these findings 
through various models. They proposed a constant Seyfert-like component that is thought to be superimposed over a variable non-thermal beamed jet power law. The presence for a soft excess has been established, yet its origin remains ambiguous. It has been suggested by \cite{2004MNRAS.349...57P} that this soft excess may arise from the inverse Comptonization of accretion disk photons 
by secondary thermal electrons at a temperature of $\sim$ 350 eV.\\ 
\\
Furthermore, \cite{2008A&A...479..365P} developed a model for the Seyfert-like component where an emsemble of coronal loops emit 
radiation across the soft to hard X-ray spectrum, thereby providing a plausible explanation for the soft excess. An anti-correlation between optical emissions and radio emissions during the period from 2008 to 2015 has been interpreted by \cite{2020MNRAS.497.2066F} as resulting from an ejection from the jet, which leads to an increase in radio synchrotron emissions following the infall of the inner region of the accretion disk into the black hole, consequently causing a reduction in disk emissions.

Similar observations have been reported for other radio galaxies, such as 3C 120 \citep{2002Natur.417..625M} and 
\cite{2011ApJ...734...43C}, which establish a direct link between the black hole and the jet. The characteristics of radio loud AGNs, including flatter X-ray spectra and weaker Fe K$\alpha$ lines, have been discussed by \cite{2020ApJ...901..111K}. This phenomenon may be attributed to strong jet contamination affecting the hard X-ray spectra \citep{2002NewAR..46..221G}. The X-ray emission from the jet component tend to be flatter. Also, it provides little contribution to the emission of the accretion disk and equatorial material. In radio galaxies, potential mechanisms that could account for the weak X-ray reflection include a highly ionized inner accretion disk \citep{2002MNRAS.332L..45B}; variations in geometry of inner disk \citep{2000ApJ...537..654E}, central accretion flow obscured by the corona or the jet \citep{2009ApJ...700.1473S} and outflowing corona which yields weaker reflection due to higher bulk outflowing velocity  \citep{2014ApJ...794...62B, 2017ApJ...835..226K}.  \\
\\
The X-ray band 0.4-78 keV consists of the soft X-ray excess below $\sim$2 keV, Compton reflection hump peaking at $\sim$20 keV, and 
jet power-law continuum components. For testing the relativistic reflection for the first time to all the observations, we modeled the 
broadband using relxill plus an extra power law model which described well all the spectral components \citep[e.g.,]{2014ApJ...782...76G,
 2014MNRAS.444L.100D, 2022MNRAS.514.3965D}. The origin of soft 
excess, which is still not clear, is likely due to the high density of the inner accretion disk \citep{2024MNRAS.534..608M}.
 Along with the soft X-ray excess, the Compton reflection 
hump is also well described. The radius-dependent power law index is found to be very high (see Table 4) and the inner radius of the 
accretion is observed to be $R_{in}\sim 2~r_g$. The soft X-ray emission is well fited  by the blurred reflection model. The inner 
radius and break radius both are very close to the event horizon and hence the emission obtained is very much smeared due to the 
strong relativistic effects suggested by the index parameter with a very high value of $\sim 9$. The emission lines in the soft X-ray 
band are blurred due to the strong relativistic effects \citep{2022MNRAS.514.3965D, 2020A&A...644A.132U, 2013MNRAS.435.1287P}. Here, 
the deviations from the power law continuum could be possibly due to the very high accretion disk density. These parameters suggest 
that the soft X-ray emission is generated in the vicinity of the SMBH. The ionization parameter is very low or almost close to zero. 
This refers to the low ionized or neutral accretion disk surface. Such a disk reflects the high energy emission with characteristic 
X-ray emission of Fe-K$\alpha$ line. This emission generated close to the black hole is strongly affected by the GR/SR effects. 
Such effects broaden the emission lines too much to be detected with the current observations. This could be a reason that we are 
unable to detect any broad Fe-K emission line in any of our observations. In X-ray modeling, the jet continuum is dominated at higher 
energies and this component is always present in all observations.  \\
\\ 
 In addition, from the results obtained from the broad-band SED modeling, we found the magnetic field strength to 
be $B \sim 0.4$ gauss at the subparsec scale. According to \citet{1977MNRAS.179..433B}, the magnetic field strength 
depends on the observed 2-10 keV luminosity, mass of the black hole and the size of the emitting region. We used 2-10 keV 
luminosity to be $3.8\times10^{45}$ erg/s, mass of the blackhole as $\sim~8\times10^{8}~M_{\odot}$ and the size of 
emitting region to be $10~r_g$ (coronal size suggested by \citep{2009Natur.459..540F, 2021Natur.595..657W} for estimating 
the magnetic field strength in the vicinity of black hole in the innermost sphere of radius 10$r_{g}$). Comparing the total 
magnetic energy in the innermost region with the net equipartition kinetic energy distribution, we concluded that the temperature 
of the particles responsible for the magnetic energy is very large compared to the coronal temperature $\sim50$ keV (as obtained 
in the reflection modeling). Thus, this calculation hints possibly a different plasma responsible for disk reflection.  \\
\\
We find an inverse correlation between spectral slope, $\alpha$ versus flux which indicates that the spectra become hardened when flux increases. \cite{2002MNRAS.336..932K} found spectra became steeper with an increase in source flux which is similar to those found in Seyfert galaxies \citep{2009MNRAS.399.1597S, 2012A&A...537A..87C}. This is attributed to the emission from the accretion disc and/or thermal coronal emission. In some of the previous studies, such correlation was absent between flux and spectral 
index \citep{2004MNRAS.349...57P,2007A&A...465..147C,2008A&A...486..411S}. The harder-when-brighter trend is also found for a sample of low luminosity AGNs \citep{2009MNRAS.399..349G} and X-rays are attributed to the Comptonization of seed photons in ADAF 
(advection-dominated accretion flows). The phenomenon was elucidated by \cite{2011MNRAS.417..280S} as being the result of a hot corona that is outflowing and located above the accretion disc, which is the site of the the seed photons for Compton scattering.
 In our observations, the hardening of the spectra with flux rise at low energies (0.4--10) keV energy band can be explained using a two-component scenario i.e. AGN is bright at low flux levels softening the index and jet contribution increases with flux leading to the hardening of the spectrum. The current data makes it challenging to distinguish the contributions of jet component and AGN component to the observed variability, as these two components are not anticipated to exhibit correlation \citep{2008A&A...486..411S}.\\
\\
\section{Conclusions}
\begin{itemize}
\item The count rates in different energy bands for all five epochs show clear variation at low energies.
\item The source exhibits a harder-when-brighter nature which indicates the variations driven by the jet emission.
\item Our broadband SED modeling infers that the jet emission arises from the edge of BLR where EC contribution from BLR and 
dusty torus is significant. 
\item The broadband SED modeling fails to explain the multiband X-ray spectra during low states due to spectral flattening at 
low energy X-rays, which can most probably be associated with the reprocessed emission and are further indicated based on the 
presence of soft excess and weak Compton hump.
\item For the first time, we tested here to model these residuals using the reflection model. The soft X-ray excess and the weak Compton hump are well described with the relativistic model such as relxillCp along with the power law model for jet emission.
\item  The average plasma temperature is found to be $\sim$15 keV which is consistent with the CompTT disk model with a jet component fitted by \cite{2015ApJ...812...14M} 
\item According to the equipartition energy distribution, the coronal temperature is likely found to be different compared to 
that of the particle population responsible for the jet emission.

\end{itemize} 
\begin{table}
\caption{Observation log of 3C 273 of XMM--Newton (pn data--X) and NuSTAR data (N). }
\vspace*{0.1in}
\noindent
\begin{tabular}{ccccc}\hline 
Observation&Obsid&Duration& F$_{var}$\\ 
date and time& &(in ks)&\\\hline
2015-07-13 (X1)&0414191101& 72.4&1.04$\pm$0.14\\
2016-06-26 (X2)&0414191201& 67.2&$0.64 \pm0.12$\\
2017-06-26 (X3)&0414191301& 67.0&$0.75 \pm0.13$\\
2018-07-04 (X4)&0414191401& 78.0&-\\
2019-07-02 (X5)&0810820101& 69.4&-\\\hline\hline
2015-07-13 (N1)& 10002020003&49.41&-\\
2016-06-26 (N2)& 10202020002&35.42&-\\
2017-06-26 (N3)& 10302020002&35.40& $2.48 \pm0.87$\\
2018-07-04 (N4)& 10402020006&40.32&-\\
2019-07-02 (N5)& 10502620002&49.41& $3.05 \pm0.95$\\ \hline
\end{tabular} \\
\noindent
\end{table}

 \begin{table}
 \caption{Observation log of 3C 273 with XMM-Newton OM (Imaging Mode)}
 \textwidth=6.0in
 \scriptsize
 \setlength{\tabcolsep}{0.035in}
 %\vspace*{0.2in}
 \noindent
 \begin{tabular}{ccccc} \hline \hline
 
 Date of Observation   &Filter &Count  &Magnitude$^a$  &Flux$^b$  \\
  dd.mm.yyyy           &       &Rate    &           &  \\ \hline
 13.07.2015            &V      &131.35$\pm$0.357 &12.67$\pm$0.003 &3.28$\pm$0.009   \\
                       &U      &303.87$\pm$0.297 &12.05$\pm$0.001 &5.89$\pm$0.006\\
                       &B      &254.71$\pm$0.276 &13.25$\pm$0.001 &3.18$\pm$0.003 \\
                       &UVW1   &150.17$\pm$0.352 &11.76$\pm$0.002 &7.24$\pm$0.017 \\
                       &UVM2   &41.96$\pm$0.134  &11.71$\pm$0.003 &9.27$\pm$0.030 \\
                       &UVW2   &18.60$\pm$0.086  &11.69$\pm$0.005 &10.60$\pm$0.049  \\ \hline
 26.06.2016            &V      &134.03$\pm$0.344 &12.64$\pm$0.003 &3.35$\pm$0.009   \\
                       &U      &349.24$\pm$0.317 &11.90$\pm$0.001 &6.77$\pm$0.006 \\
                       &B      &293.63$\pm$0.294 &13.10$\pm$0.001 &3.67$\pm$0.004 \\
                       &UVW1   &180.07$\pm$0.433 &11.56$\pm$0.003 &8.67$\pm$0.021 \\
                       &UVM2   &52.31$\pm$0.154  &11.48$\pm$0.003 &11.56$\pm$0.034 \\
                       &UVW2   &23.83$\pm$0.098  &11.42$\pm$0.004 &13.58$\pm$0.056 \\ \hline
 26.06.2017            &B      &256.47$\pm$0.276 &13.24$\pm$0.001 &3.21$\pm$0.003 \\
                       &UVW1   &149.70$\pm$0.343 &11.77$\pm$0.002 &7.22$\pm$0.017 \\
                       &UVM2   &41.62$\pm$0.134  &11.72$\pm$0.003 &9.20$\pm$0.030 \\
                       &UVW2   &18.78$\pm$0.056  &11.68$\pm$0.003 &10.70$\pm$0.032 \\ \hline
 04.07.2018            &V      &136.73$\pm$0.356 &12.62$\pm$0.003 &3.42$\pm$0.009  \\
                       &U      &297.23$\pm$0.291 &12.08$\pm$0.001 &5.77$\pm$0.006 \\
                       &B      &255.62$\pm$0.269 &13.25$\pm$0.001 &3.19$\pm$0.003 \\
                       &UVW1   &152.98$\pm$0.365 &11.74$\pm$0.003 &7.37$\pm$0.018 \\
                       &UVM2   &43.63$\pm$0.139  &11.67$\pm$0.003 &9.64$\pm$0.031 \\
                       &UVW2   &19.68$\pm$0.090  &11.63$\pm$0.005 &11.22$\pm$0.051 \\\hline
 02.07.2019            &V      &136.74$\pm$0.381 &12.62$\pm$0.003 &3.42$\pm$0.009 \\
                       &U      &296.47$\pm$0.290 &12.08$\pm$0.001 &5.75$\pm$0.006 \\
                       &B      &255.50$\pm$0.270 &13.25$\pm$0.001 &3.19$\pm$0.003 \\
                       &UVW1   &154.38$\pm$0.262 &11.73$\pm$0.002 &7.44$\pm$0.013 \\
                       &UVM2   &43.62$\pm$0.098  &11.67$\pm$0.002 &9.64$\pm$0.022 \\
                       &UVW2   &19.70$\pm$0.064  &11.63$\pm$0.003 &11.23$\pm$0.036 \\ \hline
 \end{tabular}
 
 $^a$ Instrumental Magnitude. \\
 $^b$ Flux in units of 10$^{-14}$ erg cm$^{-2}$ s$^{-1}$ A$^{-1}$. \\
 \end{table}

\begin{table}
 \caption{Fitting results of Fe K $\alpha$ emission }
 \textwidth=6.0in
 \textheight=9.0in
\setlength{\tabcolsep}{0.0010in}

 \vspace*{0.1in}
 \noindent
 \begin{tabular}{ccc} \hline
Observation                   &$\sigma$     &EW (eV) \\
 date and time                &(in keV)         &(eV)               \\ \hline
 2015-07-13         &$0.27^{+0.21}_{-0.17}$    &$20.37^{+1.17}_{-0.40}$   \\
 2016-06-26         &$0.26^{+0.51}_{-0.21}$    &$7.25^{+0.03}_{-0.69}$ \\
 2017-06-26         &$0.20^{+0.30}_{-0.17}$    &$8.93^{+0.30}_{-0.56}$    \\
 2018-07-04         &$0.41^{+6.27}_{-0.35}$    &$19.50^{+0.64}_{-1.24}$    \\
 2019-07-02         &$0.17^{+1.34}_{-0.11}$    &$9.33^{+0.91}_{-0.21}$   \\ \hline

\end{tabular} \\
     \noindent
\end{table}

\begin{figure*}
\centering
\includegraphics[width=7cm, angle=0]{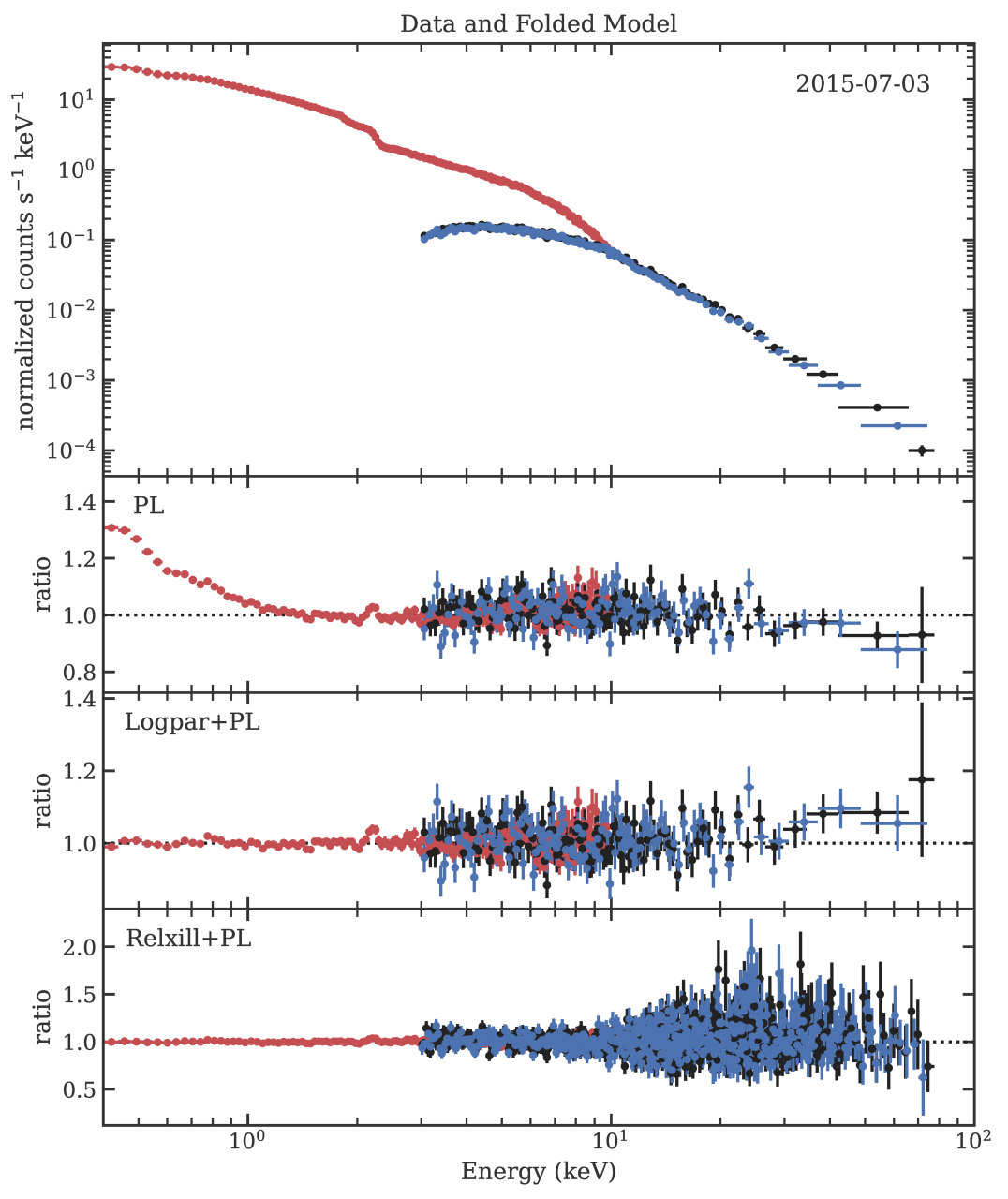}
\includegraphics[width=7cm, angle=0]{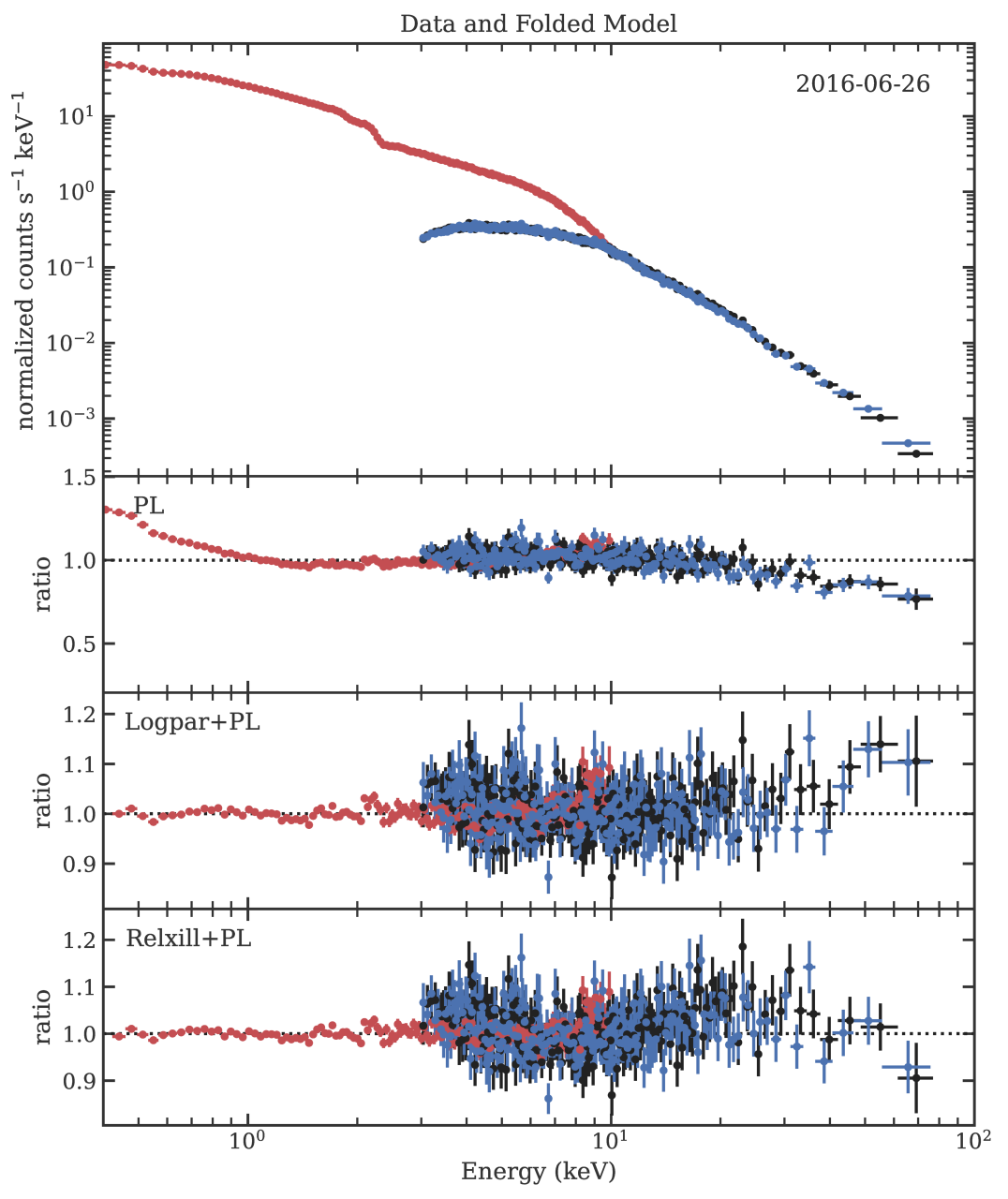}
\includegraphics[width=7cm, angle=0]{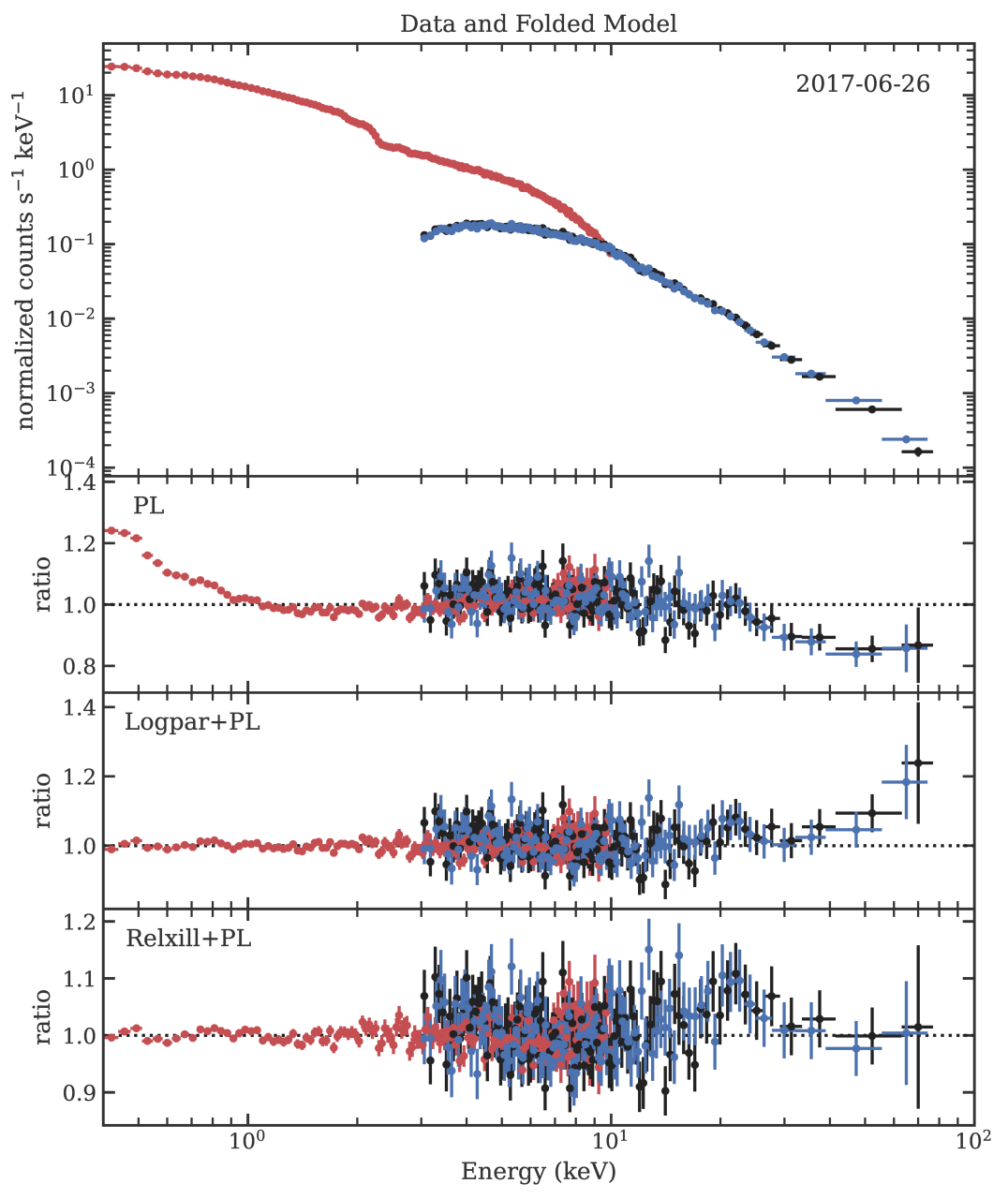}
\includegraphics[width=7cm, angle=0]{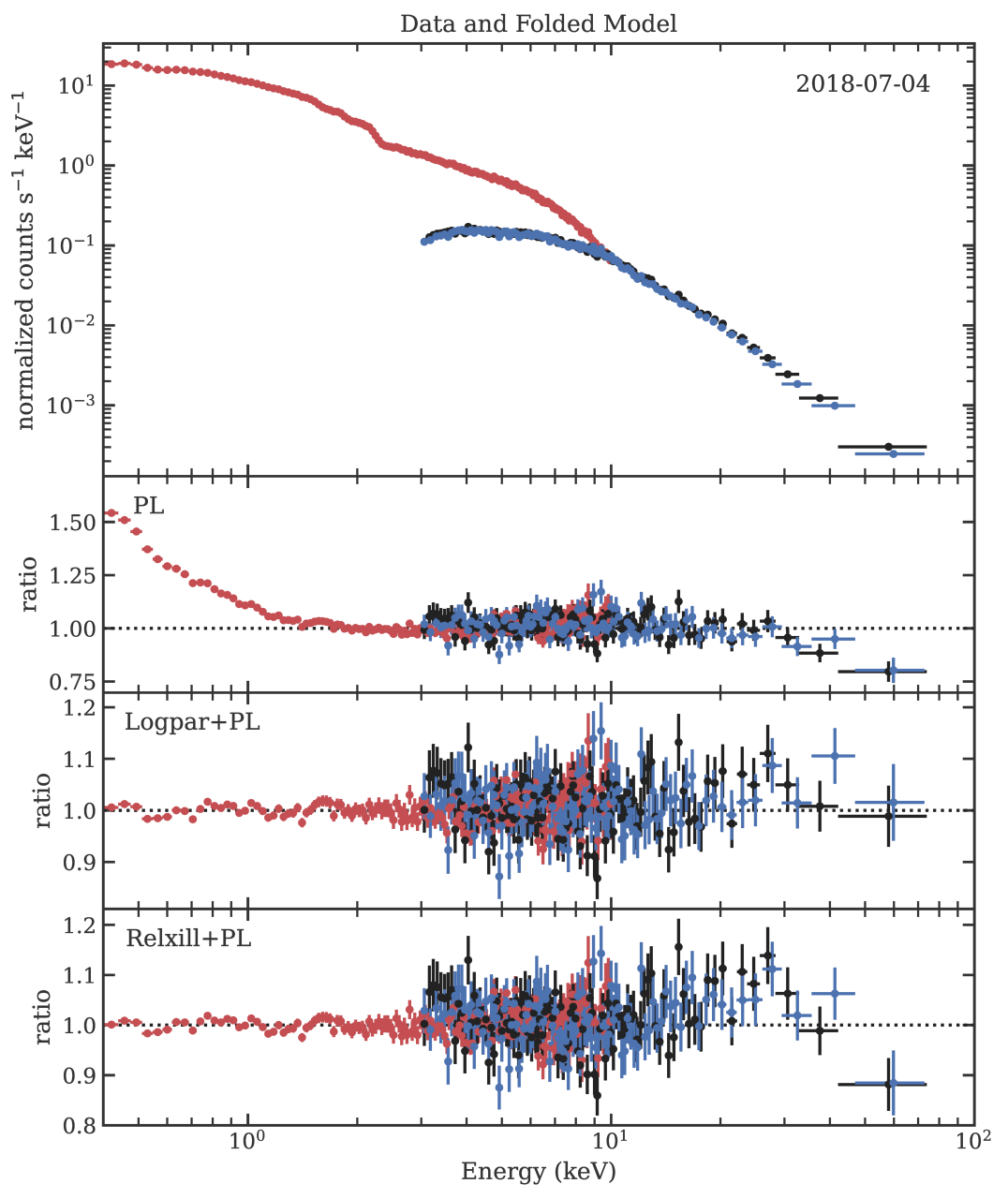}
\includegraphics[width=7cm, angle=0]{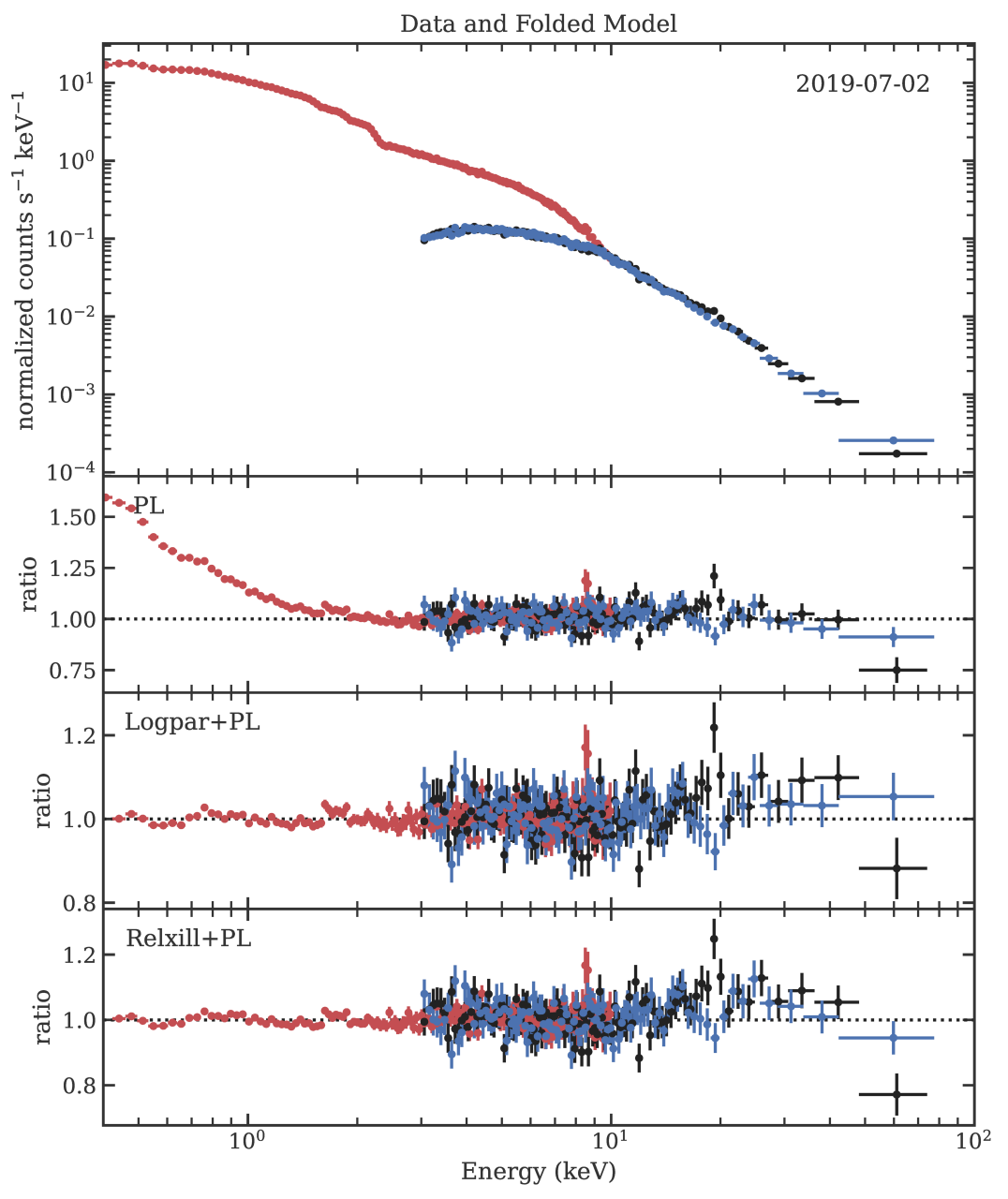}
\caption{The spectral fitting of five observations of 3C 273 in (0.3--80) KeV band. }
 \end{figure*}

\begin{table*}
%\textwidth=6.4in
%\textheight=10.0in
%\scriptsize
\centering
\caption {Fitting results of powerlaw, logpar plus powerlaw and relxill plus powerlaw for 3C 273}
\begin{tabular}{cccccc}\hline
Parameters &\textbf{July-2015}& \textbf{June-2016}&\textbf{June-2017}& \textbf{July-2018}&\textbf{July-2019}\\\hline
&&Powerlaw&&\\\hline
Photon Index&1.69$^{+0.01}_{-0.01}$&1.560$^{+0.004}_{-0.004}$&1.60$^{+0.01}_{-0.01}$&1.67$^{+0.01}_{-0.01}$&1.69$^{+0.01}_{-0.01}$\\
Norm (10$^{-2}$)&1.38$^{+0.01}_{-0.01}$&2.45$^{+0.01}_{-0.01}$&1.35$^{+0.01}_{-0.01}$&1.32$^{+0.01}_{-0.01}$&1.19$^{+0.01}_{-0.01}$\\
Stat ($\chi^{2}/\nu$)&1245.5/1185&2193.7/1473&1405.0/1171&1256.2/1138&1141.3/1156\\
\hline
&&Logpar + Powerlaw&&\\\hline
$\alpha$&$1.27^{+0.08}_{-0.09}$&$1.02^{+0.05}_{-0.05}$&$1.08^{+0.08}_{-0.09}$&$1.16^{+0.08}_{-0.09}$&$1.18^{+0.10}_{-0.10}$\\
$\beta$&$0.18^{+0.03}_{-0.03}$&$0.25^{+0.02}_{-0.02}$&$0.24^{+0.04}_{-0.03}$&$0.23^{+0.04}_{-0.03}$&$0.20^{+0.04}_{-0.04}$\\
Norm (10$^{-3}$)&$7.49^{+,1.12}_{-0.1.10}$&$11.9^{+1.1}_{-1.1}$&$6.5^{+1.0}_{-1.0}$&$6.5^{+1.0}_{-1.0}$&$5.4^{+0.9}_{-0.9}$\\
Photon Index&$2.49^{+0.10}_{-0.09}$&$2.38^{+0.06}_{-0.06}$&$2.34^{+0.10}_{-0.08}$&$2.57^{+0.08}_{-0.07}$&$2.52^{+0.08}_{-0.07}$\\
Norm (10$^{-3}$)&$7.11^{+1.11}_{-1.13}$&$13.19^{+1.1}_{-1.1}$&$7.3^{+1.0}_{-1.0}$&$8.3^{+0.9}_{-0.9}$&$8.3^{+0.9}_{-1.0}$\\
Stat ($\chi^{2}/\nu$)&1263.9/1224&1837.1/1512&1286.7/1210&1224.8/1177&1217.6/1195\\
\hline
%&&&Pexrav&&\\\hline
%Photon Index&2.906$^{+0.151}_{-0.084}$&2.796$^{+0.021}_{-0.073}$&2.652$^{+0.062}_{-0.068}$&2.792$^{+0.000}_{-0.000}$&2.690$^{+0.000}_{-0.000}$\\
%foldE (keV)&40.256$^{+132.476}_{-17.800}$&39.583$^{+43.986}_{-6.421}$&18.546$^{+14.719}_{-2.312}$&1.203e+05$^{+0.0003}_{-0.0002}$&9.999e+05$^{+0.002}_{-0.002}$\\
%rel-refl&10.7632$^{+7.815}_{-5.656}$&7.116$^{+1.981}_{-1.569}$&8.152$^{+3.304}_{-4.456}$&2.297$^{+0.000}_{-0.000}$&2.059$^{+0.000}_{-0.000}$\\
%Norm (10$^{-3}$)&3.238$^{+0.4}_{-0.7}$&6.77913$^{+0.8}_{-0.2}$&4.008$^{+0.5}_{-0.4}$&5.842$^{+0.000}_{-0.000}$&6.441$^{+0.000}_{-0.000}$\\
%Stat ($\chi^{2}/\nu$)&1271.3/1222&1893.2/1510&1291.1/1208&1250.2/1175&1326.1/1193\\\hline
&&Relxill + Powerlaw&&\\\hline
Photon Index & 1.29$_{-0.12}^{+0.11}$&1.09$^{+0.05}_{-0.1}$         &1.16$^{+0.11}_{-0.21}$       &1.19$_{-0.09}^{+0.08}$  &1.18$^{+0.01}_{-0.01}$\\
Norm (10$^{-3}$) & 2.4$\pm{1.0}$&2.4$^{+0.6}_{-0.8}$ &1.8$\pm 1.0$              &1.5$\pm{0.6}$ &2.7$^{+0.4}_{-0.5}$\\
Spin (a)&0.998(f)&0.998(f)&0.998(f)&0.998(f)&0.998(f)\\
$R_{in}$ ($R_g$)& 1.6$^{+0.2}_{-0.1}$&$<1.5$              &1.4$\pm 0.1$              &1.4$^{p}$&1.7$^p$\\
$R_{br}$ ($R_g$) &3.1$^{+0.3}_{-0.1}$&4.0$^{+0.4}_{-0.3}$ &12.2$_{-6.0}^{+53.5}$     &1.4${t}$&6.8$^{+16.7}_{-0.1}$\\
Index1 & $>9.0$&7.6$^{+0.2}_{-0.3}$                  &$>8.5$                    &4.4$\pm 0.3$&7.9$^{+0.9}_{-0.2}$\\
Index2 & 3$^{f}$&3$^f$&3$^f$&3$^f$&1.40$_{-0.04}^{+0.08}$\\
gamma & 1.97$\pm 0.03$&1.80$^{+0.01}_{-0.01}$        &1.85$^{+0.02}_{-0.03}$    &1.99$\pm 0.02$ &2.000$\pm 0.003$\\
log$x_{i}$ & 0.8$\pm0.2$&1.29$^{+0.2}_{-0.02}$            &0.8$^{+0.2}_{-0.3}$       &0(p)&0.9$^{+0.1}_{-0.2}$\\
logN (cm$^{-3}$)&18.5$\pm 0.3$&18.9$^{+0.2}_{-0.1}$&18.9$\pm 0.2$            &19.4$\pm 0.2$ &19.0$^{+0.02}_{-0.1}$\\
refl$_{frac}$ & 1$^f$&1$^f$&1$^f$&1$^f$&1$^f$\\
kT$_{e}$ (keV)&11.1$^{+3.4}_{-2.1}$& 15.4$^{+3.2}_{-3.0}$ &11.7$^{+3.6}_{-2.0}$ &$15.1_{-3.4}^{+4.1}$   &47.2$^{+27.7}_{-18.4}$  \\
norm (10$^{-4}$)&1.7$\pm 0.2$&4.1$\pm 0.3$           &1.9$\pm 0.3$              &1.9$\pm 0.1$      &1.9$\pm 0.1$\\
Stat ($\chi^{2}$/$\nu$)& 1231.6 / 1217 & 1779.5 / 1506 & 1263.2 / 1203 & 1252.4/1186 & 1260.1 / 1188 \\
flux (0.3-3.0 keV)$10^{-11}$& 5.2 &9.3   &5.0  &5.4    & 5.0  \\
flux (10-40 keV)$10^{-11}$&8.3 &21.3   &10.4  &8.3    &7.1   \\
\hline
\end{tabular}\\
$a$: Spin of the black hole  \\
$R_{g}$: Inner and outer radius of the accretion disk  \\
Index1/Index2: Emissivity for the coronal model as given as $r^{-Index1}$ and $r^{-Index2}$ \\
gamma: Power law index of the incident spectrum\\
log$x_{i}$: Ionization parameter  \\
log N: density of the accretion disk   \\
K$T_{e}$: Temperature of plasma \\
\end{table*}
%\textwidth=6.4in

%\textheight=10.0in

%\scriptsize

\begin{figure*}
\centering
\includegraphics[width=6cm, angle=-90]{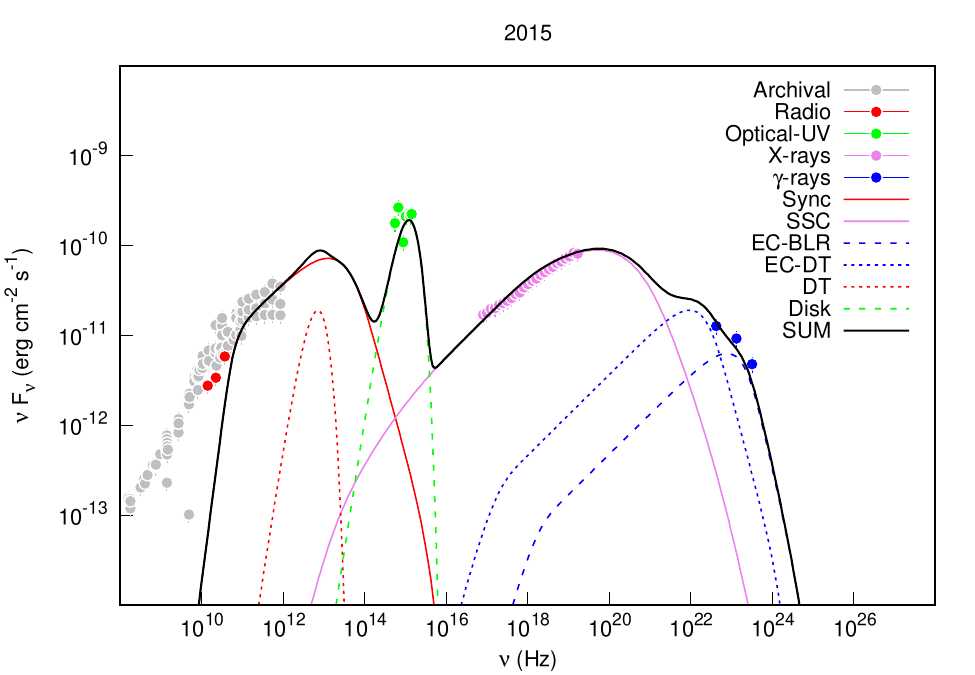}
\includegraphics[width=6cm, angle=-90]{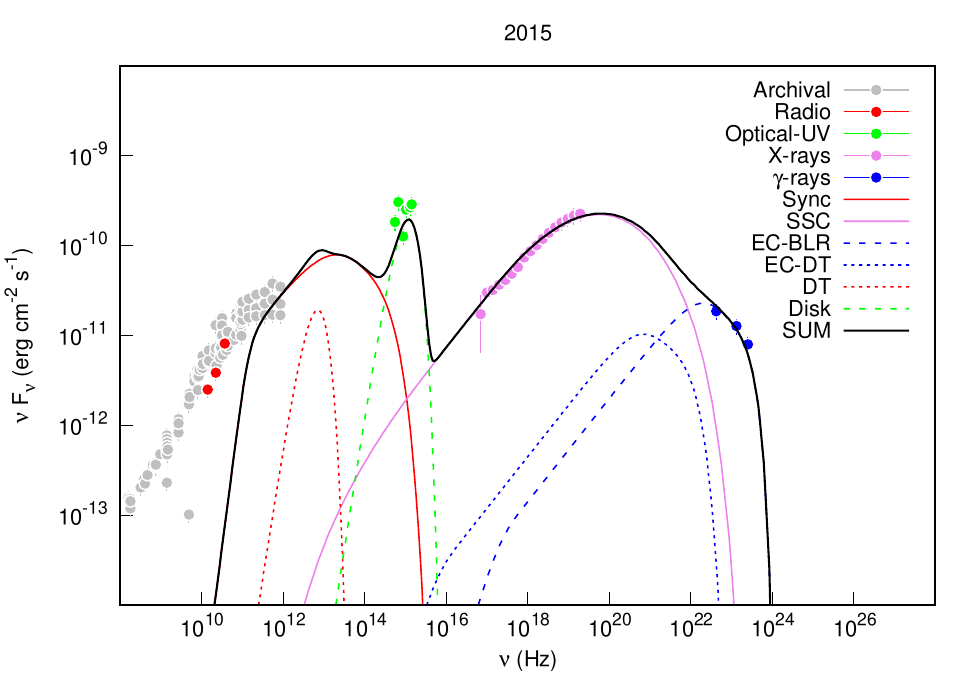}
\includegraphics[width=6cm, angle=-90]{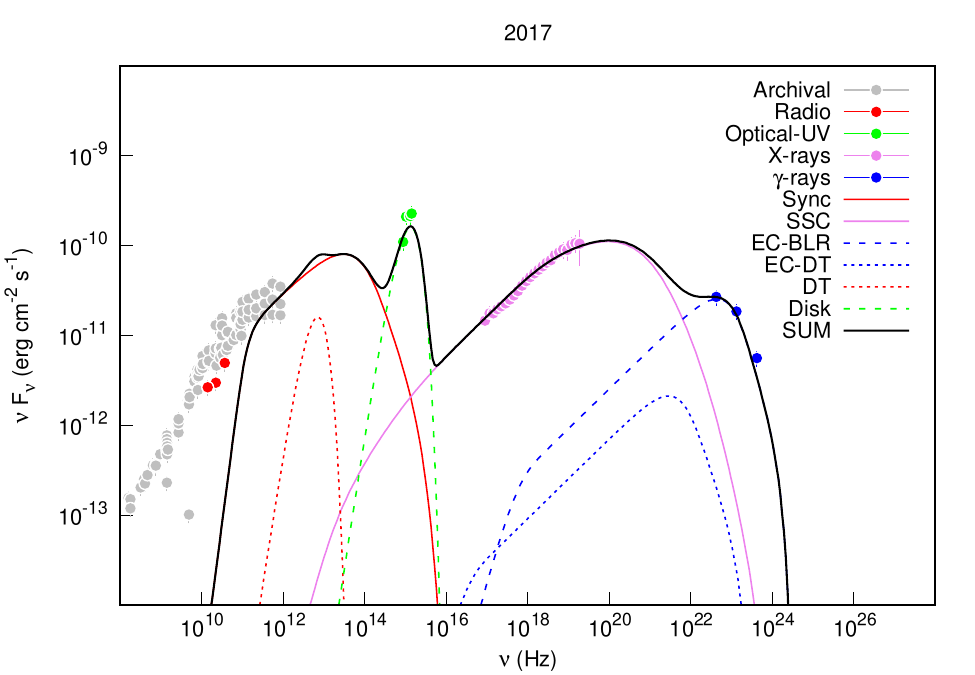}
\includegraphics[width=6cm, angle=-90]{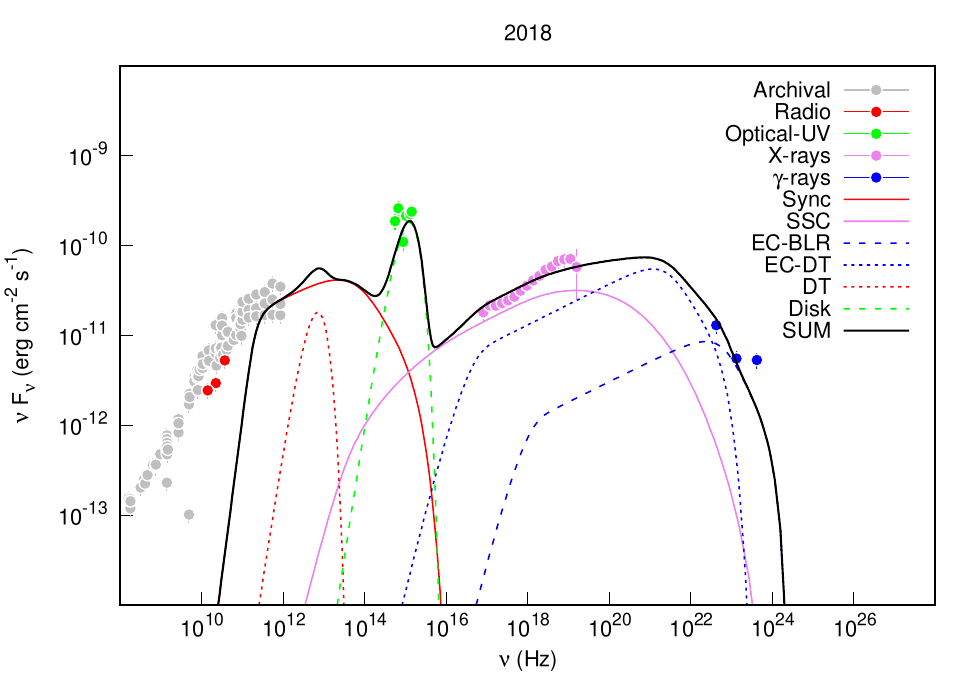}
\includegraphics[width=6cm, angle=-90]{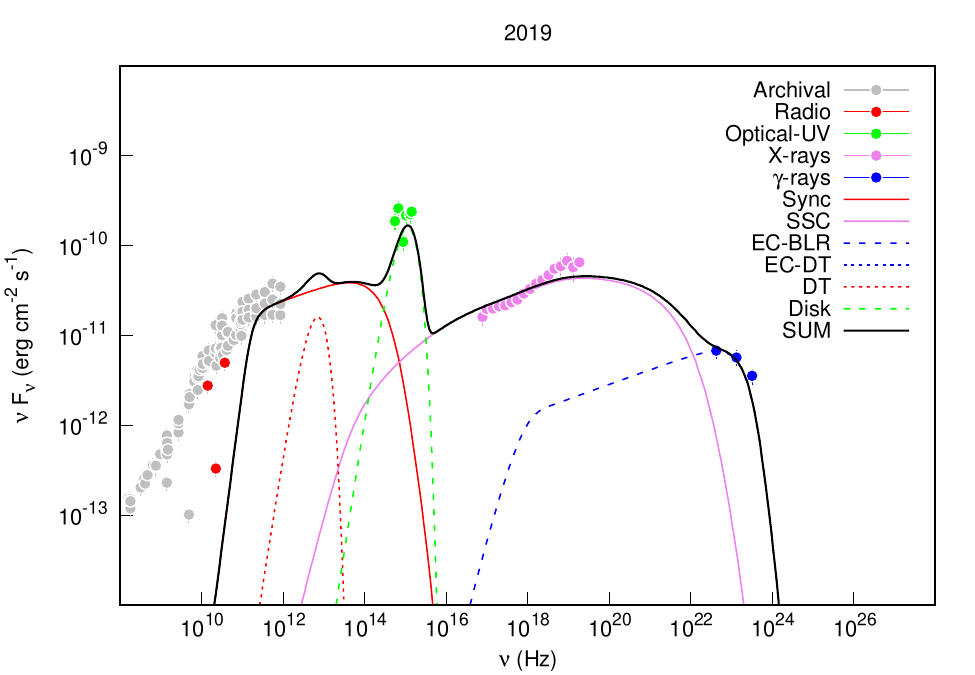}

\caption{One-zone synchrotron+IC modelling of SED of 3C 273 during all epochs.}
 \end{figure*}

\begin{table}
\caption{Physical parameters of one-zone model for 3C 273.}
.\setlength{\tabcolsep}{0.035in}
%\vspace*{0.2in}
\noindent

\begin{tabular}{cccccccccc}
\hline
Year & $R$ (cm) & $B$ (G) & $\delta$ & $\gamma_{min}$ & $\gamma_{max}$ & $\gamma_b$  & $L_{D}$(erg s$^{-1}$) & $T_D$ (K) \\
\hline
2015 & 2.0E17 & 0.09 & 8.8 & 8.0 & 2.0E4 & 3.0E3 & 1.9E46 & 1.7E4 \\
2016 & 8.0E16 & 0.08 & 4.0 & 3.0 & 5.0E3 & 1.0E3 & 1.8E46 & 1.5E4 \\
2017 & 9.4E16 & 0.48 & 5.7 & 6.0 & 1.0E4 & 2.0E3 & 1.6E46 & 1.9E4 \\
2018 & 6.4E16 & 0.84 & 5.5 & 7.0 & 8.0E3 & 2.0E3 & 1.8E46 & 1.7E4 \\
2019 & 8.0E16 & 0.45 & 6.1 & 8.0 & 1.0E4 & 4.0E3 & 1.6E46 & 1.6E4 \\
\hline
\end{tabular}
\end{table}

\begin{acknowledgments}
 HG acknowledges the financial support from the Science and Engineering Research (SERB), India through SERB Research Scientist
 award SB/SRS/2022-23/113/PS at ARIES, Nainital. Main Pal is thankful for financial support from IUCAA, Pune through its Visiting 
Associate Programme. AP acknowledges support from the Polish Funding Agency National Science Centre, project 2017/26/A/ST9/00756 
(MAESTRO 9). LC acknowledges support by the National Natural Science Foundation of China (grant 12173066), the National SKA Program 
of China (Grant No.2022SKA0120102) and Shanghai Pilot Program for Basic Research, Chinese Academy of Science, Shanghai Branch 
(JCYJ-SHFY-2021-013).
\end{acknowledgments}

\bibliographystyle{aasjournal}

{}
\end{document}